# Imaging Exploration of Molecular Subtypes in Tongue Squamous Cell Carcinoma

Hao Pan[1], Peipei Wang[1], Yajie Chang[1], Bingyi Lu[1], Yunyan Jiang[1], Mengfan Wang[1], Xinyue Wang[1], Xinrou Yang[1], Jiyuan Zhang[1], Yu Liu[3], Andrei Velichko[2,*], Yuanjun Wang[1,*]

Affiliations:
1   School of Health Science and Engineering, University of Shanghai for Science and Technology, Shanghai 200093, China
2   Institute of Physics and Technology, Petrozavodsk State University, Petrozavodsk, Russia
3   Department of Radiology, Shanghai Ninth People's Hospital Affiliated Shanghai Jiao Tong University School of Medicine, 639 Zhizaoju Road, Huangpu District, 200011, Shanghai, China

## Abstract

Tongue squamous cell carcinoma (TSCC) is an aggressive malignancy with marked biological heterogeneity and variable clinical outcomes. Although molecular profiling has improved understanding of TSCC heterogeneity, its clinical use remains constrained by invasive tissue sampling and limited representation of whole-tumor spatial complexity. Meanwhile, most radiomics studies in TSCC have focused on downstream clinical endpoints, and whether imaging can non-invasively reflect intrinsic molecular subtypes remains unclear. In this study, an integrated transcriptomic-radiomics framework was used to investigate the relationship between preoperative imaging phenotypes and molecular subtypes in TSCC. Transcriptomic data from 60 TSCC cases in The Cancer Genome Atlas were analyzed using unsupervised consensus clustering, followed by differential expression and functional enrichment analyses. Matched preoperative imaging data from The Cancer Imaging Archive were manually annotated for primary tumor regions, and radiomic features were extracted using PyRadiomics; group differences were assessed with the U-test. Two stable molecular subtypes, C1 and C2, were identified. Their biological differences were mainly associated with squamous epithelial differentiation, inflammatory signaling, and lipid metabolism, with C2 showing greater enrichment of immune-related pathways. In addition, 10 radiomic features differed significantly between the two subtypes, mainly wavelet-derived texture features from gray-level size zone, dependence, co-occurrence, and run length matrices ($P=0.00202$-$0.0162$). These findings support the potential of radiomics as a non-invasive approach for characterizing molecular heterogeneity in TSCC and provide an initial radiogenomic framework for biologically informed preoperative assessment.

# Introduction

Tongue squamous cell carcinoma (TSCC) is a highly aggressive malignancy characterized by rapid local invasion, a high propensity for cervical lymph node metastasis, frequent locoregional recurrence, and substantial variation in survival outcomes. Even among patients with apparently similar clinicopathologic characteristics, treatment response and long-term prognosis can differ considerably, indicating that conventional staging and histopathologic assessment do not fully capture the intrinsic biological diversity of TSCC[1–3].

Recent molecular studies have increasingly shown that TSCC is not a biologically uniform disease. Spatial transcriptomic analysis has revealed marked transcriptional reprogramming across different tissue regions in tongue cancer, while multi-omics studies integrating genome, methylome, and transcriptome data have identified molecular events associated with prognosis and tumor progression[4–6]. In parallel, mechanistic work has demonstrated that local tissue components can actively participate in TSCC evolution; notably, tongue muscle cells can transdifferentiate into cancer-associated fibroblasts in response to TSCC, highlighting the importance of stromal remodeling in shaping the tumor ecosystem[5]. A recent systematic review of omics-derived prognostic biomarkers in TSCC further emphasized that a wide range of candidate molecular signatures has been reported, but that the overall biomarker landscape remains heterogeneous and insufficiently translated into clinically usable stratification tools[7]. Together, these observations support the view that TSCC comprises biologically distinct subgroups with differences in differentiation state, stromal interaction, inflammatory signaling, and therapeutic vulnerability.

Although molecular profiling has substantially advanced the biological understanding of TSCC, its clinical use remains constrained by tissue dependence. Biopsy-based assays are invasive, difficult to repeat longitudinally, and inherently limited in their ability to represent the full spatial complexity of the entire tumor. This limitation is particularly relevant in TSCC, where invasive fronts, stromal reactions, and biologically distinct intratumoral regions may coexist within the same lesion[4,5]. Consequently, there is a strong need for a non-invasive method capable of assessing whole-tumor heterogeneity before treatment and translating molecular complexity into clinically actionable information.

Radiomics has emerged as a promising approach for extracting quantitative, high-dimensional phenotypic information from routine medical images. In TSCC, MRI- and CT-based radiomics models have already shown potential for preoperative prediction of cervical lymph node metastasis, pathological stage, locoregional recurrence, and overall survival[8–15]. Recent studies have reported encouraging results from multimodal MRI radiomics, deep learning–radiomics integration, machine-learning–based MRI models, and fusion strategies that incorporate clinical features or intratumoral heterogeneity descriptors[9–16]. These findings suggest that imaging contains biologically relevant information beyond gross morphology and that radiomics may help capture clinically meaningful tumor heterogeneity in TSCC.

However, the current radiomics literature in TSCC remains predominantly endpoint-oriented. Most published models focus on predicting nodal metastasis, recurrence, or survival, whereas far fewer studies address the upstream biological basis of these imaging phenotypes. Moreover, multicenter validation has shown that previously reported MRI radiomics models do not always retain robust performance across datasets, underscoring persistent concerns regarding reproducibility and generalizability[12]. Thus, a key unresolved question is not only whether radiomics can predict clinical outcomes in TSCC, but whether it can identify biologically meaningful tumor subtypes in a non-invasive and interpretable manner. This question is particularly relevant because subtype-specific differences in squamous differentiation, cell-adhesion architecture, metabolic activity, inflammatory state, and stromal remodeling are all biologically plausible determinants of tumor margin characteristics, internal heterogeneity, and texture complexity on imaging. This is an inference drawn from the convergence of TSCC molecular studies and endpoint-focused imaging studies, rather than a claim already firmly established by direct TSCC radiogenomic evidence.

Against this background, the present study was designed to investigate whether radiomic features derived from preoperative imaging can discriminate molecular subtypes of TSCC and thereby serve as non-invasive surrogates of the tumor's intrinsic biological state. By linking imaging phenotypes to molecular stratification, this study aims to move beyond conventional prognostic modeling and establish a more biologically interpretable framework for preoperative TSCC assessment. Such a strategy may improve risk stratification, support individualized therapeutic planning, and facilitate the clinical translation of TSCC molecular heterogeneity into imaging-based precision management.

## Methods

**2.1 Study Design**

This study was designed to investigate the molecular heterogeneity of tongue squamous cell carcinoma (TSCC) and its corresponding imaging phenotypic characteristics through an integrated transcriptomic–radiomics framework. First, TSCC cases with the primary tumor located in the tongue were identified from The Cancer Genome Atlas (TCGA), and transcriptomic data were used to perform unsupervised consensus clustering for molecular subtype discovery. Based on the identified subtypes, differential expression analysis was subsequently conducted, followed by Gene Ontology (GO) and Kyoto Encyclopedia of Genes and Genomes (KEGG) enrichment analyses, in order to characterize the biological differences between subtypes.

Corresponding preoperative imaging data were then retrieved from The Cancer Imaging Archive (TCIA) for matched cases. The primary tumor region of interest (ROI) was manually delineated for radiomics feature extraction, and an integrative imaging–molecular analysis was performed to evaluate whether radiomic features could distinguish molecular subtypes and reflect underlying tumor biological heterogeneity. The overall workflow of the study is presented in Figure 1.

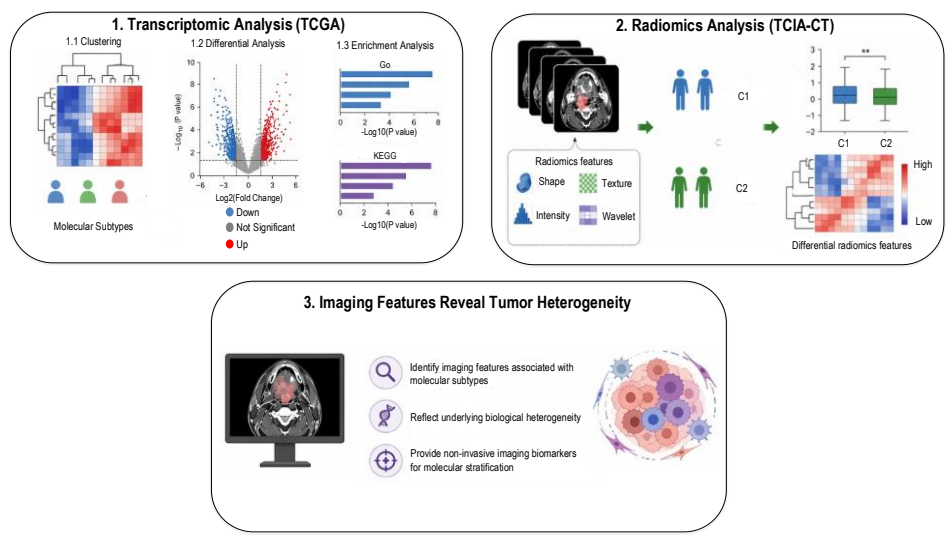

Figure 1. Overall workflow of the study. TSCC cases were selected from TCGA for transcriptomic analysis and molecular subtype identification using unsupervised consensus clustering. Differential expression and functional enrichment analyses were then performed to characterize subtype-related biological differences. Corresponding preoperative imaging data from TCIA were subsequently used for ROI annotation and radiomics feature extraction. Finally, an imaging–molecular integrative analysis was conducted to explore whether radiomic features could reflect molecular heterogeneity in TSCC.

### 2.2 Data Acquisition and Sample Selection

Transcriptomic data and corresponding clinical information were obtained from the TCGA-HNSC public database. The cases were selected based on the criterion of the primary tumor site being located in the tongue, yielding 60 TSCC samples for subsequent analysis. Imaging data were sourced from TCIA, and corresponding preoperative imaging data were retrieved using case identifiers to match the TCGA samples for imaging preprocessing, ROI annotation, and imaging phenotype analysis.

### 2.3 Transcriptome Data Preprocessing

The selected TSCC transcriptomic data were preprocessed and quality-controlled. First, the expression matrix and corresponding clinical data were organized, and samples with significant data loss or non-compliance with inclusion criteria were excluded. The remaining expression data were then normalized to reduce the bias from sequencing depth and technical variation. The preprocessed expression matrix was used for unsupervised clustering, differential expression analysis, and functional enrichment analysis.

### 2.4 Unsupervised Consensus Clustering for Molecular Subtype Identification

To identify potential molecular subtypes of TSCC, simple consensus clustering was performed on the transcriptomic data. Euclidean distance was used to measure the similarity between samples, with a resampling ratio set to 0.8, and 80% of the samples were selected for each iteration, repeated 1000 times[17]. The clustering results were visualized using a consensus matrix, which helps ensure the stability and consistency of the clustering. Ultimately, 2 molecular subtypes were identified and designated as C1 and C2 for further analysis.

### 2.5 Differential Expression Analysis

Differential expression analysis was performed using the limma package[18]. The threshold for identifying differentially expressed genes was set to $|log2FC| > 1$ and adjusted $P < 0.05$. The differential expression results were visualized using a heatmap to illustrate the gene expression patterns across the two subtypes. The selected differential genes were further analyzed through GO and KEGG enrichment analysis.

### 2.6 Functional Enrichment Analysis

To clarify the biological basis of the differences between molecular subtypes, GO and KEGG enrichment analyses were performed on the differentially expressed genes. The results indicated that the different subtypes were enriched in biological processes related to lipid metabolism, inflammatory responses, and squamous epithelial differentiation. The C2 subtype showed greater enrichment in immune-related pathways, such as cytokine interactions and the IL-17 signaling pathway.

### 2.7 Imaging Annotation

The preoperative imaging data obtained from TCIA were processed, organized, and standardized. ROI annotation was performed manually around the primary tumor, focusing on tumor boundaries and internal heterogeneity. The ROI was delineated using ITK-SNAP software[19]. Radiomics features were extracted using the PyRadiomics package[20], which includes morphological, texture, and gray-level co-occurrence matrix (GLCM) features.

**2.8 Integrative Analysis**

An integrative analysis framework was built based on the identified molecular subtypes and annotated imaging data. The aim was to explore whether different molecular subtypes exhibited stable imaging phenotype features and how the biological differences of these subtypes were reflected in the imaging characteristics.

**2.9 Statistical Analysis**

All bioinformatics and radiomics analyses were performed in standard R and Python computing environments. The unsupervised clustering results were visualized using a consensus matrix. The differential expression results were presented as a heatmap, and the functional enrichment analysis results were displayed using standard pathway analysis plots. Statistical analysis was performed using U-test to compare group differences, with a significance level set at $P < 0.05$. All statistical analyses were performed using two-sided tests, with appropriate multiple comparison corrections applied where necessary.

# Result

**3.1 Unsupervised consensus clustering identified two stable molecular subtypes**

Based on the transcriptomic profiles of 60 TSCC samples from the TCGA cohort, unsupervised consensus clustering was performed to identify molecular subtypes. The consensus matrix showed that the samples could be stably separated into two subgroups, designated as C1 and C2 (Figure2A). Two well-defined high-consensus blocks were observed along the diagonal of the matrix, whereas the between-cluster regions showed relatively low consensus, indicating good intracluster stability and intercluster separation across repeated resampling. The hierarchical clustering dendrogram further supported the two-cluster solution, suggesting that TSCC is not a homogeneous disease entity and can be divided into at least two relatively stable molecular subtypes.

**3.2 Distinct transcriptomic patterns were observed between the C1 and C2 subtypes**

To further characterize the molecular differences between the two subtypes, differential expression analysis was performed and a heatmap of representative differentially expressed genes was generated (Figure2B). The heatmap showed a clear separation in global expression patterns between C1 and C2. A subset of genes was relatively upregulated in C1 and downregulated in C2, whereas another subset showed the opposite trend, indicating distinct transcriptional characteristics between the two subtypes.

As shown in the heatmap, these differentially expressed genes were able to discriminate C1 from C2 clearly, suggesting that the identified clustering pattern was not only stable but also associated with substantial molecular expression differences. Overall, these findings further support the presence of reproducible molecular heterogeneity within TSCC.

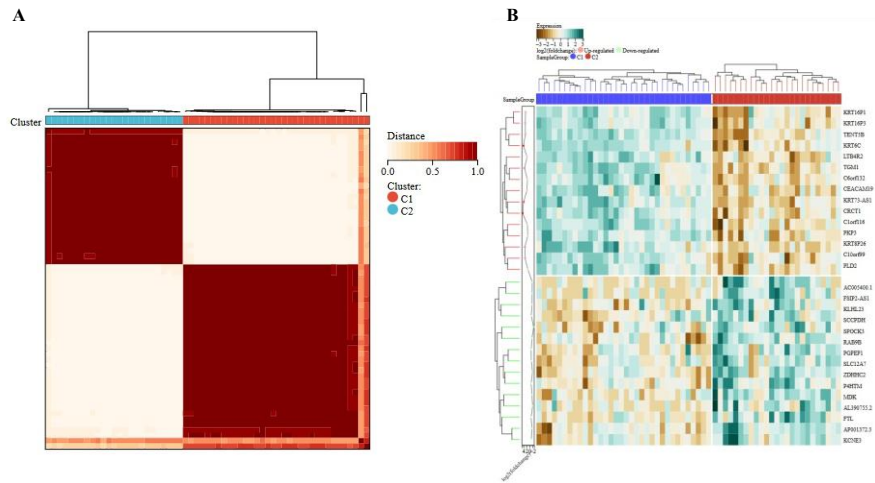

Figure 2. Unsupervised consensus clustering identified two molecular subtypes in TSCC and distinct differential expression patterns between the subtypes.(A) Consensus matrix derived from unsupervised consensus clustering based on transcriptomic data.(B) Heatmap of representative differentially expressed genes between the C1 and C2 subtypes.

### 3.3 Enrichment analysis of differentially expressed genes

KEGG and GO enrichment analyses were performed for the differentially expressed genes between C1 and C2. The results showed that these genes were mainly associated with metabolic reprogramming, inflammation-related responses, and squamous epithelial differentiation and structural maintenance. KEGG analysis revealed that the differentially expressed genes were mainly enriched in the Estrogen signaling pathway, Staphylococcus aureus infection, Arachidonic acid metabolism, Retinol metabolism, Chemical carcinogenesis, Ether lipid metabolism, Drug metabolism-cytochrome P450, Linoleic acid metabolism, and alpha-Linolenic acid metabolism (Figure3A). Among these, several lipid metabolism-related pathways, including arachidonic acid metabolism, linoleic acid metabolism, alpha-linolenic acid metabolism, and ether lipid metabolism, were prominently enriched, suggesting substantial metabolic differences between the two subtypes. GO enrichment analysis further showed that the differentially expressed genes were mainly involved in biological processes such as epidermis development, skin development, epidermal cell differentiation, keratinization, epidermal keratinization, intermediate filament organization, and peptide cross-linking. In terms of cellular components, these genes were mainly enriched in cornified envelope, keratin filament, intermediate filament, desmosome, plasma membrane, apical plasma membrane, and collagen-containing extracellular matrix. In terms of molecular functions, the major enriched categories included structural constituent of skin epidermis, peptidase inhibitor activity, endopeptidase inhibitor activity, endopeptidase regulator activity, lipase activity, metal ion transmembrane transporter activity, glycosaminoglycan binding, and monocarboxylic acid binding (Figure3B). Together, these findings indicate that the overall differences between the two subtypes are mainly related to squamous epithelial differentiation, cell adhesion and keratin-associated structures, as well as lipid metabolism and peptidase regulation.

### 3.4 Enrichment analysis of upregulated differentially expressed genes

Separate enrichment analysis of the upregulated differentially expressed genes showed that the main KEGG pathways included Cytokine-cytokine receptor interaction, Staphylococcus aureus infection, Estrogen signaling pathway, IL-17 signaling pathway, Arachidonic acid metabolism, Serotonergic synapse, Ether lipid metabolism, Retinol metabolism, Linoleic acid metabolism, and alpha-

Linolenic acid metabolism (Figure3C). Among these, cytokine-cytokine receptor interaction, IL-17 signaling pathway, and Staphylococcus aureus infection were among the leading enriched terms, suggesting that the upregulated genes were more strongly associated with inflammatory and immune-related activation. At the same time, arachidonic acid metabolism, linoleic acid metabolism, alpha-linolenic acid metabolism, and ether lipid metabolism were also significantly enriched, indicating that lipid-associated inflammatory mediator processes may also play an important role in this gene set.

GO analysis showed that the upregulated genes were mainly enriched in epidermis development, keratinocyte differentiation, skin development, keratinization, epidermal cell differentiation, intermediate filament cytoskeleton organization, intermediate filament-based process, and establishment of skin barrier at the biological process level. At the cellular component level, the major enriched terms included cornified envelope, intermediate filament, keratin filament, intermediate filament cytoskeleton, desmosome, apical plasma membrane, and apical part of cell. At the molecular function level, the enriched categories mainly included structural constituent of skin epidermis, serine-type endopeptidase inhibitor activity, serine-type peptidase inhibitor activity, endopeptidase inhibitor activity, peptidase inhibitor activity, and endopeptidase regulator activity (Figure3D). These results suggest that the upregulated genes were mainly associated with enhanced squamous epithelial differentiation, keratin-related structural organization, and peptidase inhibitory regulation.

### 3.5 Enrichment analysis of downregulated differentially expressed genes

KEGG enrichment analysis of the downregulated genes showed that they were mainly enriched in Metabolic pathways, Chemical carcinogenesis, Metabolism of xenobiotics by cytochrome P450, Insulin secretion, Drug metabolism-cytochrome P450, Drug metabolism-other enzymes, Salivary secretion, Dilated cardiomyopathy, Steroid hormone biosynthesis, and Tyrosine metabolism (Figure3E). Among these, Metabolic pathways showed the highest GeneRatio, and cytochrome P450-related xenobiotic metabolism, drug metabolism, and chemical carcinogenesis pathways were also prominently enriched, suggesting that the downregulated genes were mainly associated with reduced metabolic and biotransformation functions.

GO analysis showed that the downregulated genes were mainly enriched in regulation of animal organ morphogenesis, synapse assembly, regulation of developmental growth, morphogenesis of a branching structure, developmental growth involved in morphogenesis, regulation of morphogenesis of an epithelium, response to atropine, cell fate specification, respiratory system development, and kidney mesenchyme development at the biological process level. At the cellular component level, the enriched terms mainly included collagen-containing extracellular matrix, synaptic cleft, apicolateral plasma membrane, GABA-ergic synapse, synaptic membrane, sarcolemma, potassium channel complex, presynaptic membrane, and sarcoplasmic reticulum. At the molecular function level, the major enriched categories included carboxylic acid binding, organic acid binding, glycosaminoglycan binding, growth factor activity, monocarboxylic acid binding, sulfur compound binding, potassium ion transmembrane transporter activity, delayed rectifier potassium channel activity, potassium channel activity, and heparin binding (Figure3F). Overall, the downregulated genes were more closely related to metabolic conversion, ion transport, and several developmental and cell communication-related functions.

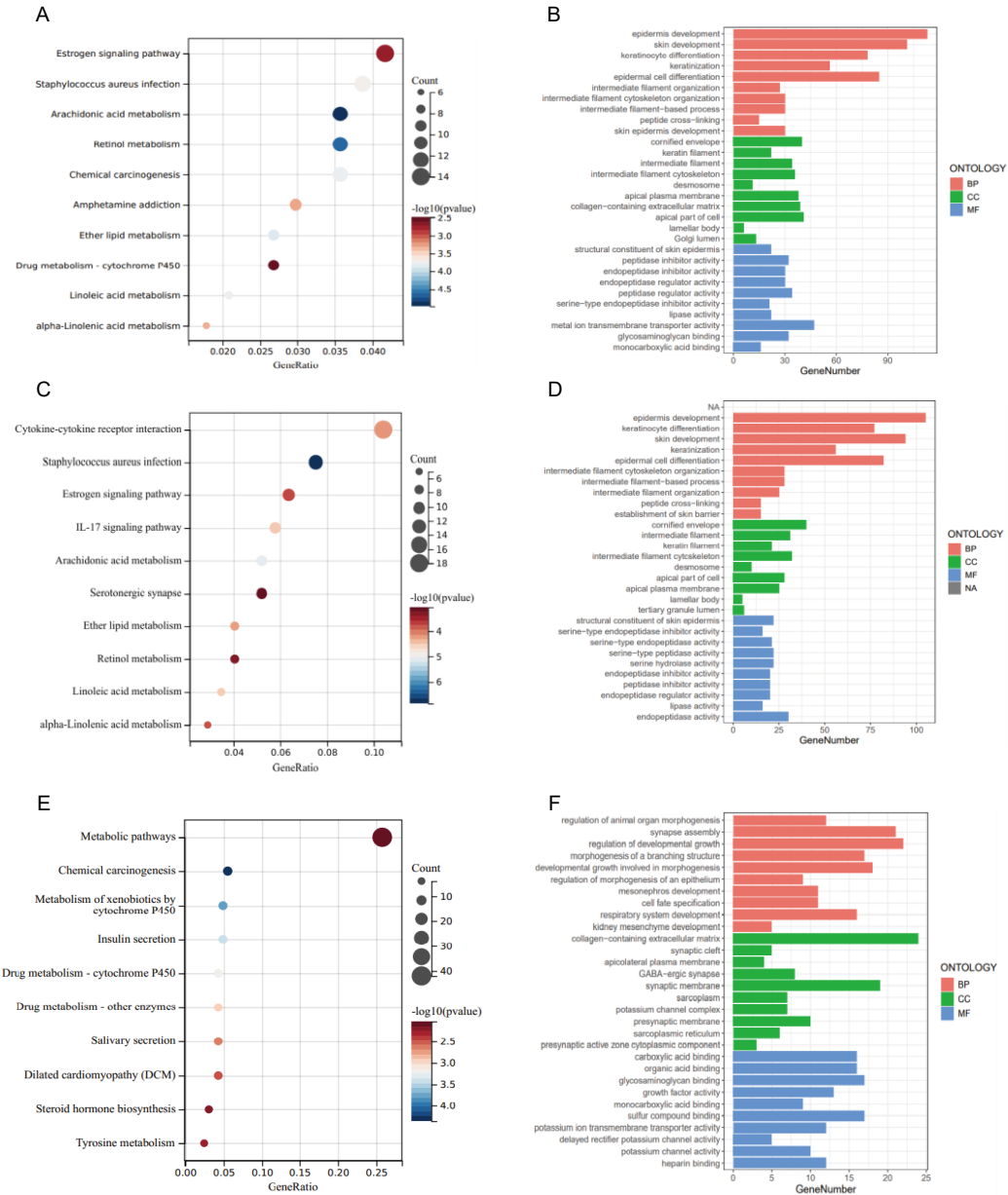

Figure 3.Enrichment analyses of differentially expressed genes, upregulated genes, and downregulated genes between the C1 and C2 subtypes.(A)KEGG enrichment analysis of differentially expressed genes.(B)GO enrichment analysis of differentially expressed genes.(C)KEGG enrichment analysis of upregulated differentially expressed genes.(D)GO enrichment analysis of upregulated differentially expressed genes.(E)KEGG enrichment analysis of downregulated differentially expressed genes.(F)GO enrichment analysis of downregulated differentially expressed genes.

**3.6Differences in radiomics features between the C1 and C2 subtypes**

To compare imaging phenotypes between the C1 and C2 molecular subtypes, candidate radiomics features were analyzed between the two groups. The results showed that 10 candidate radiomics features were significantly different between C1 and C2 (Figure4). These features were mainly derived from the gray-level size zone matrix (GLSZM), gray-level dependence matrix (GLDM), gray-level co-occurrence matrix (GLCM), and gray-level run length matrix (GLRLM), and most of them were higher-order texture features extracted after wavelet transformation.

Specifically, the significantly different features included wavelet-HHH_glszm_SizeZoneNonUniformity (P=0.00202), wavelet-HHH_glszm_SmallAreaHighGrayLevelEmphasis (P=0.00583), wavelet-HHH_glszm_SmallAreaEmphasis (P=0.00657), wavelet-LHH_glszm_ZoneEntropy (P=0.0091), wavelet-LHH_glszm_ZonePercentage (P=0.0093), wavelet-LHH_glszm_GrayLevelNonUniformity (P=0.0101), original_gldm_DependenceNonUniformityNormalized (P=0.013), log-sigma-5-mm-3D_glrlm_ShortRunEmphasis (P=0.013), original_glcm_ClusterShade (P=0.0144), and wavelet-LHH_glszm_SizeZoneNonUniformity (P=0.0162).

Overall, several wavelet-based GLSZM texture features showed clear distributional separation between C1 and C2, suggesting differences between the two subtypes in zone-size distribution, non-uniformity, small-area-related patterns, and textural complexity. In addition, GLDM-, GLCM-, and GLRLM-derived features also differed between the two groups, further indicating that C1 and C2 exhibited distinct imaging texture organization and gray-level distribution patterns. Taken together, these findings suggest that different molecular subtypes of TSCC may present radiomics phenotypes that can be captured by quantitative imaging features.

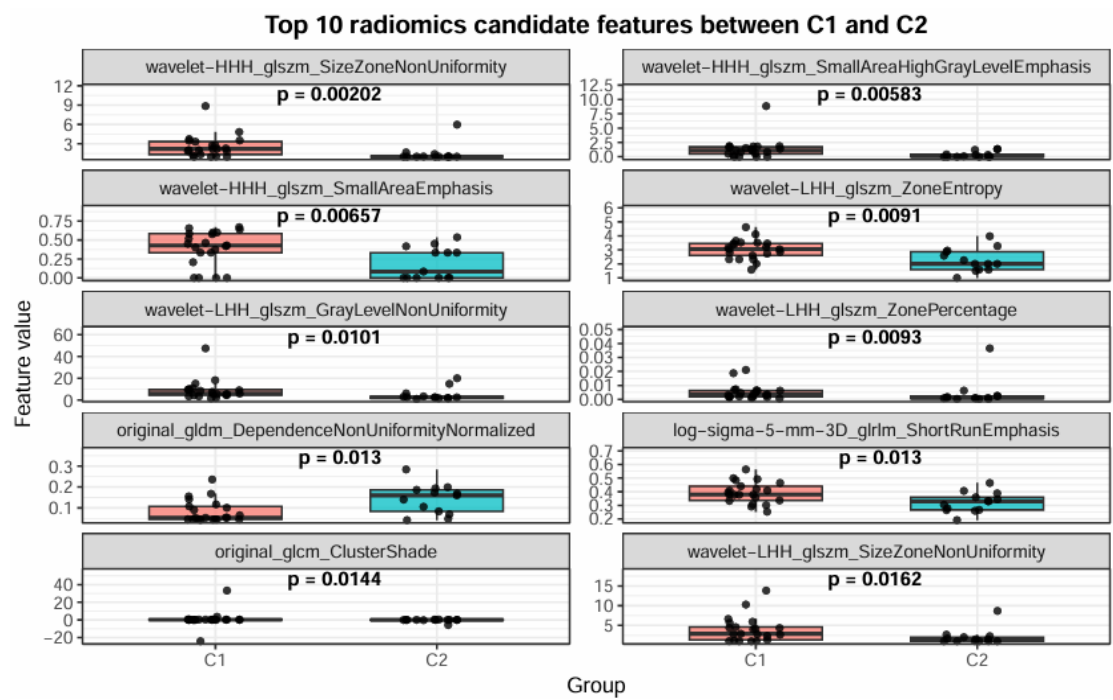

Figure 4.Comparison of differential radiomics features between the C1 and C2 subtypes

# Discussion

This study was designed to determine whether preoperative radiomic features can distinguish molecular subtypes of tongue squamous cell carcinoma (TSCC) and thus serve as non-invasive surrogates of the tumor's intrinsic biological state. Two relatively stable transcriptome-defined subtypes were identified, and their divergence was mainly characterized by differences in squamous differentiation, inflammatory signaling, and lipid-metabolic reprogramming, which were further accompanied by distinct texture-based imaging phenotypes. Taken together, these findings suggest that at least part of the biological heterogeneity of TSCC may already be encoded in routine preoperative imaging, thereby providing a basis for a more biologically interpretable framework of radiologic stratification.

The identification of two stable molecular subtypes further supports the view that TSCC is not a biologically homogeneous disease. This interpretation is consistent with recent transcriptomic and spatial studies showing that oral and tongue squamous carcinomas harbor substantial intratumoral and regional heterogeneity, with coordinated variation across epithelial programs, stromal states, and immune composition[4–7,21,24,25]. In the present study, the clear separation of transcriptomic patterns between C1 and C2 indicates that the consensus clusters were not merely statistical partitions, but likely reflected distinct dominant biological programs. Given that spatial transcriptomic work has already demonstrated marked regional transcriptional reprogramming in tongue cancer, it is plausible that the subtypes identified here capture whole-tumor tendencies emerging from heterogeneous but coordinated intralesional states, rather than isolated local fluctuations[4,21,25].

The enrichment results suggest that the biological differences between the two subtypes were organized around three interconnected axes: squamous epithelial differentiation and keratin-associated structural maintenance, inflammatory or immune-related activation, and lipid metabolism with broader biotransformation capacity. First, the repeated enrichment of epidermal development, keratinization, keratin filaments, desmosomes, and cornified-envelope-related terms indicates that subtype divergence may partly reflect differences in differentiation state and epithelial structural organization. In a squamous malignancy such as TSCC, these processes are not only histologic descriptors but may also influence the spatial arrangement of tumor cells, the coherence of cell–cell adhesion, and the structural regularity of tumor tissue, all of which could contribute to imaging texture. Second, the enrichment of cytokine–cytokine receptor interaction and IL-17-related signaling suggests that one subtype may exist in a more inflammatory microenvironment. This is biologically plausible in light of recent evidence that inflammatory and stromal signaling are deeply integrated in TSCC progression, including CAF-related remodeling and immune crosstalk [5,22,26–29]. Third, the enrichment of arachidonic acid, linoleic acid, alpha-linolenic acid, ether lipid metabolism, and cytochrome P450-associated pathways implies that subtype differences also extend to metabolic adaptation and biotransformation functions. This finding aligns with the broader view that TSCC progression is shaped not only by proliferative programs, but also by shifts in metabolic ecology and stromal support[6,7,26,27].

These molecular differences provide a coherent interpretive context for the radiomics results. The subtype-associated imaging differences were concentrated in gray-level size zone matrix (GLSZM), gray-level dependence matrix (GLDM), gray-level co-occurrence matrix (GLCM), and gray-level run length matrix (GLRLM) features, with a predominance of higher-order wavelet-derived textures. Compared with gross morphological descriptors, these features are more likely to reflect internal signal complexity, local non-uniformity, repetitive microstructural organization, and heterogeneity in gray-level distribution. The fact that differences were particularly evident in features related to zone-size distribution, non-uniformity, small-area patterns, and entropy suggests that the divergence between C1 and C2 may be expressed less through macroscopic shape and more through tissue-level complexity. A restrained interpretation is therefore that differences in differentiation-related architecture, inflammatory composition, and stromal or metabolic remodeling may collectively reshape the spatial organization of intratumoral signals and thus become detectable as texture phenotypes. Importantly, the present study does not establish a one-to-one mechanistic correspondence between a specific pathway and an individual radiomic feature. Rather, it shows that molecular subtype structure and quantitative imaging texture are meaningfully linked at the phenotype level.

This distinction is important when positioning the present work relative to the current TSCC radiomics literature. Most previous studies in TSCC have focused on clinically downstream endpoints, including lymph node metastasis, pathological stage, locoregional recurrence, or survival [9–16,21,29,30]. These studies collectively support the clinical relevance of radiomics, but they often leave the upstream biological basis of imaging phenotypes insufficiently explained. By contrast, the current study shifts the analytical anchor from outcome prediction to molecular stratification. In doing so, it addresses a more fundamental question: not only whether radiomics can predict aggressive behavior, but also what biological variation those image-derived features may be reflecting. This repositioning is important because the translational value of radiomics will likely depend not only on statistical performance, but also on interpretability and reproducibility. Recent work in oral squamous cell carcinoma (OSCC) has further shown that MRI radiomics can be associated with immune gene-expression signatures and tumor-infiltrating lymphocyte-related phenotypes, supporting the broader concept that imaging features may act as non-invasive proxies for microenvironmental states[22,23,30].

The present findings are also broadly concordant with recent work emphasizing the importance of the TSCC and OSCC microenvironment. Spatial and single-cell studies have shown that fibroblasts, monocytic populations, T-cell states, and tumor-border immune organization are not passively associated with disease progression, but actively shape invasion, differentiation loss, and clinical outcome[24,25,31–34]. In parallel, experimental studies have demonstrated that CAFs can remodel the extracellular matrix, alter matrix stiffness, and promote invasion-related phenotypes, while tongue-specific work has highlighted the dynamic conversion of local tissue components into tumor-supportive stromal cells[5,26–29]. Within this context, it is biologically credible that subtype-specific differences in stromal remodeling and inflammatory signaling may contribute to the texture complexity observed on imaging. Thus, the present results should be understood not as isolated radiomic associations, but as part of an emerging radiogenomic framework in which imaging heterogeneity reflects the composite effects of epithelial differentiation, immune activity, and stromal organization.

From an academic perspective, this study contributes to TSCC research by proposing an initial bridge between molecular subtype structure and preoperative whole-tumor imaging phenotype. Its methodological value lies in moving radiomics away from a purely endpoint-oriented black-box role toward a more interpretable intermediate phenotypic layer. This is particularly relevant in TSCC, where biopsy-based molecular assessment is inherently limited by invasiveness and incomplete spatial sampling. If validated further, such a framework could help explain why image-derived heterogeneity sometimes carries prognostic information and, conversely, why radiomics models may fail to generalize when their biological anchors are weak or inconsistent[12,30]. From a translational perspective, the current findings suggest that radiomics may eventually serve as a complementary approach for identifying biologically distinct TSCC subsets before treatment, thereby assisting preoperative stratification and potentially informing individualized planning. At present, however, the most appropriate interpretation remains that radiomics is a candidate adjunct for biological stratification rather than a substitute for molecular profiling.

Several future directions arise directly from these results. First, the subtype-associated radiomic patterns should be validated in larger independent cohorts with clearly reported imaging-matching sample sizes and external testing. Second, future studies should integrate radiomics with clinicopathologic variables, conventional imaging descriptors, and, where available, transcriptomic or immune signatures to assess whether biological subtype information provides additive value beyond standard endpoint models. Third, the mechanistic basis of the observed texture differences should be examined more directly through spatial transcriptomics, regional pathology correlation, multiplex tissue imaging, or immunohistochemistry, especially at the invasive front. Fourth, technical robustness should be strengthened through observer reproducibility testing, feature stability analysis, and standardized image-processing pipelines, given the recognized reproducibility challenges in head and neck radiomics[12,19,20,30].

Several limitations should also be acknowledged. First, this was an exploratory public-database-based study, and the effective imaging–molecular sample size was limited, which constrains statistical stability and may amplify feature-level variability. Second, the radiomic analysis relied on manually delineated regions of interest; although this is reasonable for capturing tumor boundaries and internal heterogeneity, the absence of explicitly reported interobserver reproducibility limits the assessment of segmentation robustness. Third, no external validation cohort was included, and thus the present findings should be interpreted primarily as proof of biological plausibility rather than evidence of immediate clinical utility. Fourth, the mechanistic interpretation remained indirect because it was inferred from enrichment patterns rather than validated by spatial pathology, protein-level assays, or functional experiments. Finally, the present work did not directly link the identified molecular subtypes to treatment response or survival endpoints within the same analytic framework, which limits immediate clinical extrapolation.

In conclusion, this study indicates that TSCC can be stratified into two transcriptomically distinct molecular subtypes characterized mainly by differences in squamous differentiation, inflammatory signaling, and lipid-metabolic reprogramming, and that these differences are accompanied by measurable texture-based radiomic phenotypes on preoperative imaging. More importantly, the study provides an initial radiogenomic interpretive framework for TSCC, which now requires validation in larger, standardized, and biologically integrated cohorts before it can be translated into clinically reliable preoperative stratification.

# References

1. Ghazi, N., Saghravanian, N., Anvari, K., Saghafi Khadem, S., Hoseinzadeh, M., & Barzanouni, R. (2025). Evaluation of survival rate in patients with tongue squamous cell carcinoma: a retrospective single-center study. *BMC Oral Health*, *25*(1), 658.
2. Damazo, B. J., Punjabi, N. A., Liu, Y. F., & Inman, J. C. (2024). Histopathologic predictors of recurrence and survival in early T stage oral tongue squamous cell carcinoma. *Frontiers in oral health*, *5*, 1426709.
3. Kwak, J. H., Ji, Y. B., Song, C. M., Lee, Y. J., Park, H. J., & Tae, K. (2025). Long-term oncologic outcomes and prognostic factors related to recurrences in pathologic Stage I/II early oral tongue cancer. *Frontiers in surgery*, *12*, 1534274.
4. Patysheva, M. R., Kolegova, E. S., Khozyainova, A. A., Prostakishina, E. A., Korobeynikov, V. Y., Menyailo, M. E., ... & Denisov, E. V. (2024). Revealing molecular mechanisms of early-onset tongue cancer by spatial transcriptomics. *Scientific Reports*, *14*(1), 26255.
5. Lin, W., Huang, W., Mei, S., Lu, X., Zheng, J., Wang, H., ... & Zhang, Y. (2025). Transdifferentiation of tongue muscle cells into cancer-associated fibroblasts in response to tongue squamous cell carcinoma. *Nature Communications*, *16*(1), 6753.
6. Liang, L., Li, Y., Ying, B., Huang, X., Liao, S., Yang, J., & Liao, G. (2023). Mutation-associated transcripts reconstruct the prognostic features of oral tongue squamous cell carcinoma. *International Journal of Oral Science*, *15*(1), 1.
7. Astreidis, I., Kostidis, I., Malousi, A., Paraskevopoulos, K., Andreadis, D., Vahtsevanos, K., & Vizirianakis, I. (2026). Omics-Derived Prognostic Biomarkers in Tongue Squamous Cell Carcinoma: A Systematic Review with Risk-of-Bias Appraisal and Translational Prioritization. *Current Issues in Molecular Biology*, *48*(4), 389.
8. Lai, J., Yang, H., & Chen, J. (2024). Predicting radiotherapy efficacy and prognosis in tongue squamous cell carcinoma through an in-depth analysis of a radiosensitivity gene signature. *Frontiers in Oncology*, *14*, 1334747.
9. Wang, D., He, X., Huang, C., Li, W., Li, H., Huang, C., & Hu, C. (2024). Magnetic resonance imaging-based radiomics and deep learning models for predicting lymph node metastasis of squamous cell carcinoma of the tongue. *Oral Surgery, Oral Medicine, Oral Pathology and Oral Radiology*, *138*(1), 214-224.
10. Liu, S., Zhang, A., Xiong, J., Su, X., Zhou, Y., Li, Y., ... & Liu, F. (2024). The application of radiomics machine learning models based on multimodal MRI with different sequence combinations in predicting cervical lymph node metastasis in oral tongue squamous cell carcinoma patients. *Head & Neck*, *46*(3), 513-527.
11. Yao, Y., Jin, X., Peng, T., Song, P., Ye, Y., Song, L., ... & An, P. (2024). A novel nomogram for predicting overall survival in patients with tongue squamous cell carcinoma using clinical features and MRI radiomics data: a pilot study. *World Journal of Surgical Oncology*, *22*(1), 227.
12. Tagliabue, M., Ruju, F., Mossinelli, C., Gaeta, A., Raimondi, S., Volpe, S., ... & Ansarin, M. (2024). The prognostic role of MRI-based radiomics in tongue carcinoma: a multicentric validation study. *La radiologia medica*, *129*(9), 1369-1381.
13. Vidiri, A., Dolcetti, V., Mazzola, F., Lucchese, S., Laganaro, F., Piludu, F., ... & Marzi, S. (2025). MRI in Oral Tongue Squamous Cell Carcinoma: A Radiomic Approach in the Local

Recurrence Evaluation. *Current Oncology*, *32*(2), 116.
14. Vidiri, A., Marzi, S., Piludu, F., Lucchese, S., Dolcetti, V., Polito, E., ... & Covello, R. (2023). Magnetic resonance imaging-based prediction models for tumor stage and cervical lymph node metastasis of tongue squamous cell carcinoma. *Computational and Structural Biotechnology Journal*, *21*, 4277-4287.
15. Li, W., Li, Y., Wang, L., Yang, M., Iikubo, M., Huang, N., ... & Liu, Y. (2025). Evaluating fusion models for predicting occult lymph node metastasis in tongue squamous cell carcinoma. *European Radiology*, *35*(9), 5228-5238.
16. Li, Y., Huang, N., Wang, L., Xiao, H., Chen, W., Xing, Y., ... & Li, W. (2026). An interpretable machine learning model using SHapley Additive exPlanations for preoperative cervical lymph node metastasis risk stratification in tongue squamous cell carcinoma: a multicenter study. BMC Oral Health, 26(1), 185.
17. Wilkerson, Matthew D., and D. Neil Hayes. "ConsensusClusterPlus: a class discovery tool with confidence assessments and item tracking." Bioinformatics 26.12 (2010): 1572-1573.
18. Liu, S., Wang, Z., Zhu, R., Wang, F., Cheng, Y., & Liu, Y. (2021). Three differential expression analysis methods for RNA sequencing: limma, EdgeR, DESeq2. *JoVE (Journal of Visualized Experiments)*, (175), e62528.
19. Yushkevich, P. A., Gao, Y., & Gerig, G. (2016, August). ITK-SNAP: An interactive tool for semi-automatic segmentation of multi-modality biomedical images. In *2016 38th annual international conference of the IEEE engineering in medicine and biology society (EMBC)* (pp. 3342-3345). IEEE.
20. Van Griethuysen, J. J., Fedorov, A., Parmar, C., Hosny, A., Aucoin, N., Narayan, V., ... & Aerts, H. J. (2017). Computational radiomics system to decode the radiographic phenotype. *Cancer research*, *77*(21), e104-e107.
21. Ren, J., Yang, G., Song, Y., Zhang, C., & Yuan, Y. (2024). Machine learning-based MRI radiomics for assessing the level of tumor infiltrating lymphocytes in oral tongue squamous cell carcinoma: a pilot study. *BMC Medical Imaging*, *24*(1), 33.
22. Corti, A., Lenoci, D., Corino, V. D., Mattavelli, D., Ravanelli, M., Poli, T., ... & Mainardi, L. (2025). Interplay between MRI radiomics and immune gene expression signatures in oral squamous cell carcinoma. Scientific Reports, 15(1), 12622.
23. Corti, A., De Cecco, L., Cavalieri, S., Lenoci, D., Pistore, F., Calareso, G., ... & Mainardi, L. (2023). MRI-based radiomic prognostic signature for locally advanced oral cavity squamous cell carcinoma: development, testing and comparison with genomic prognostic signatures. *Biomarker Research*, *11*(1), 69.
24. Näsiaho, J., Nissi, L., Ventelä, S., Irjala, H., & Salmi, M. (2026). Spatial single-cell analysis reveals tumor microenvironment signatures predictive of oral cavity cancer outcome. *Cell Reports Medicine*, *7*(2).
25. Sun, L., Kang, X., Wang, C., Wang, R., Yang, G., Jiang, W., ... & Sun, S. (2023). Single-cell and spatial dissection of precancerous lesions underlying the initiation process of oral squamous cell carcinoma. *Cell discovery*, *9*(1), 28.
26. Jin, W., Yu, Q., Yu, L., Zhou, T., Wang, X., Lu, W., ... & Ni, Y. (2025). HAS1high cancer associated fibroblasts located at the tumor invasion front zone promote oral squamous cell carcinoma invasion via ECM remodeling. *Journal of Experimental & Clinical Cancer Research*, *44*(1), 238.

27. Zhang, J. Y., Zhu, W. W., Wang, M. Y., Zhai, R. D., Wang, Q., Shen, W. L., & Liu, L. K. (2021). Cancer-associated fibroblasts promote oral squamous cell carcinoma progression through LOX-mediated matrix stiffness. *Journal of Translational Medicine*, *19*(1), 513.
28. Liu, X., Wang, D., Cheng, X., Ma, Y., Li, S., Deng, J., & Qiao, C. (2025). Fibroblast–immune crosstalk in oral squamous cell carcinoma: from tumor promotion to immune evasion. *Frontiers in Immunology*, *16*, 1703277.
29. Oikawa, Y., Umakoshi, M., Suzuki, K., Kudo-Asabe, Y., Miyabe, K., Koyama, K., ... & Goto, A. (2025). Prognostic significance of cancer-associated fibroblasts and tumor-associated macrophages in the tongue squamous cell carcinoma and their correlation with tumor budding. *Oral Oncology*, *165*, 107295.
30. Corti, A., Cavalieri, S., Calareso, G., Mattavelli, D., Ravanelli, M., Poli, T., ... & Mainardi, L. (2024). MRI radiomics in head and neck cancer from reproducibility to combined approaches. *Scientific Reports*, *14*(1), 9451.
31. Song, H., Lou, C., Ma, J., Gong, Q., Tian, Z., You, Y., ... & Xiao, M. (2022). Single-cell transcriptome analysis reveals changes of tumor immune microenvironment in oral squamous cell carcinoma after chemotherapy. *Frontiers in cell and developmental biology*, *10*, 914120.
32. Vidiri, A., Dolcetti, V., Mazzola, F., Lucchese, S., Laganaro, F., Piludu, F., ... & Marzi, S. (2025). MRI in Oral Tongue Squamous Cell Carcinoma: A Radiomic Approach in the Local Recurrence Evaluation. *Current Oncology*, *32*(2), 116.
33. Einhaus, J., Gaudilliere, D. K., Hedou, J., Feyaerts, D., Ozawa, M. G., Sato, M., ... & Han, X. (2023). Spatial subsetting enables integrative modeling of oral squamous cell carcinoma multiplex imaging data. *Iscience*, *26*(12).
34. Principe, S., Mejia-Guerrero, S., Ignatchenko, V., Sinha, A., Ignatchenko, A., Shi, W., ... & Kislinger, T. (2018). Proteomic analysis of cancer-associated fibroblasts reveals a paracrine role for MFAP5 in human oral tongue squamous cell carcinoma. *Journal of proteome research*, *17*(6), 2045-2059.
35. Tan, Y., Wang, Z., Xu, M., Li, B., Huang, Z., Qin, S., ... & Huang, C. (2023). Oral squamous cell carcinomas: state of the field and emerging directions. *International journal of oral science*, *15*(1), 44.